\documentclass{ab}

\usepackage{graphicx}
\usepackage[numbers]{natbib}
\usepackage{multirow}
\usepackage{mathtools}

\begin{document}

\title{Large Scale Distribution of Galaxies in The Field HS\,47.5-22. I. Data Analysis Technique}

\author{A.~A. Grokhovskaya and S. ~N. Dodonov}

\institute{Special Astrophysical Observatory, Russian Academy of Sciences, Nizhnij Arkhyz, 369167, Russia}
 \titlerunning{Large Scale Distribution of Galaxies in The Field HS\,47.5-22. I. Data Analysis Technique}

\authorrunning{Grokhovskaya}

\date{September 3,  2018/Revised: July 26, 2019}
\offprints{A. Grokhovskaya  \email{grohovskaya.a@gmail.com} }

\abstract{
We present the results of methodological works on automated
analysis of the large scale distribution of galaxies. Selecting
candidates for clusters and groups of galaxies was carried out
using two complementary methods of determining the density
contrast maps in the narrow layers of the three-dimensional large
scale distribution of galaxies: the filtering algorithm with an
adaptive core and the Voronoi tesselation. The developed
algorithms were tested on 10 data sets of the MICE model catalog;
additionally, we determined the statistical parameters of the
obtained results (completeness, sample purity, etc.). The
constructed density contrast maps were also used to determine
voids.
}

\maketitle

\section{Introduction}
\label{intro}

Galaxies are the basic blocks that make up the Universe. However,
to date, there is still no full understanding of how they form and
evolve. Observations of galaxies made it possible to create their
morphological classification~\cite{Hubble1926:Grohovskaya_n_en}, and later,
studying their physical properties lead to a more accurate bimodal
classification~\cite{Fioc1999:Grohovskaya_n_en}. The connection between bimodal
types of galaxies and their environment was first discovered in
studies of nearby clusters. Oemler~\cite{Oemler1974:Grohovskaya_n_en} and
Dressler~\cite{Dressler1980:Grohovskaya_n_en} found the so-called
``morphology--environment density'' dependence. It manifests
itself in the fact that disc galaxies with star formation are
usually located on the outskirts of galaxy clusters, whereas red
elliptical galaxies are discovered mainly in higher density
regions. Recent works based on 2dFGRS~\cite{Madgwick2003:Grohovskaya_n_en} and
SDSS~\cite{Guo2013:Grohovskaya_n_en,Guo2014:Grohovskaya_n_en} surveys have shown that the relation
between the local environment and morphology remains unchanged for
the entire range of local densities up to field galaxies.

Additionally, other physical properties of galaxies were found to
correlate with the environment density. Kauffmann et
al.~\cite{Kauffmann2004:Grohovskaya_n_en} have shown that the local density
influences the color, the H$\alpha$ line equivalent width, and the
D4000~\AA\ jump on scales of the order of 1~Mpc\,h$^{-1}$. The
authors of~\cite{Cucciati2010:Grohovskaya_n_en} suggested for a sample of
10\,000~COSMOS field galaxies (in accordance with the earlier
works~\cite{Kauffmann2004:Grohovskaya_n_en,DeLucia2004:Grohovskaya_n_en,Cucciati2006:Grohovskaya_n_en}) that the more
massive galaxies formed in the most dense regions earlier than
galaxies of lower mass, and that complex physical processes
determined by the surroundings of the lower mass galaxies
influence their evolution.

The local high density regions are determined by groups and
clusters of galaxies which are the largest gravitationally bound
objects in the Universe, whereas low density regions are
represented by voids filling up to 95\% of the volume of the
Universe~\cite{Weygaert2011:Grohovskaya_n_en}. Determining groups
and clusters of galaxies as well as voids is an important task in
modern cosmology. There are many different observational
techniques to identify groups and clusters of galaxies: by X-ray
emission of hot
gas~\cite{Romer2001:Grohovskaya_n_en,Pierre2006:Grohovskaya_n_en,Finoguenov2007:Grohovskaya_n_en},
by the Sunyaev--Zeldovich effect in the cosmic microwave
background~\cite{Carlstrom2002:Grohovskaya_n_en,Voit2005:Grohovskaya_n_en},
through cosmic shear due to weak gravitational
lensing~\cite{Weinberg2002:Grohovskaya_n_en}, galaxy overdensities
in optical, near-infrared (NIR), and mid-IR
images~\cite{Lopes2004:Grohovskaya_n_en,Koester2007:Grohovskaya_n_en}.

In this work we present methods of automated analysis of the large
scale distribution of galaxies, which are based on reconstructing
density contrast maps by a filtering algorithm with adaptive
kernel and Voronoi tesselation. The algorithms were applied to ten
data sets from lightcone $N$-body simulation from the
MICE~\cite{Carretero2017:Grohovskaya_n_en} collaboration and
showed the results corresponding to the model distribution of
density in the studied sample. Their statistical estimation has
been made. In the next paper of the cycle we use these methods to
analyze the large scale distribution of galaxies in the
HS\,47.5-22 field based on the data observed with the 1-m Schmidt
telescope of the Byurakan astrophysical observatory (BAO NAS).

The paper is structured as follows. The second section briefly
describes the data used for testing the developed algorithms, the
methods of selecting galaxy clusters and groups from the large
scale distribution are presented: constructing density contrast
maps using methods involving adaptive aperture and Voronoi
tesselation as well as criteria for the following sampling of
candidates into structures, and the basic statistical calculations
for estimating the work of the algorithms. The third section
describes the algorithm of finding voids in two dimensional layers
of the three dimensional galaxy sample cone. In the Conclusions we
list the main results. The paper uses the $\Lambda CDM$
cosmological model with parameters \mbox{$\Omega_M=0.3$},
\mbox{$\Omega_\Lambda=0.7$} and
\mbox{$H_0=70$~km\,s$^{-1}$\,Mpc$^{-1}$}.

\section{SELECTION OF GALAXY GROUPS AND CLUSTERS}

\begin{table*}[]
\captionstyle{nonumber}
\caption{{\bf Table.} Parameters of 10 samples of objects from the MICE simulation 
} \label{Tab1:Grohovskaya_n_en}
\medskip
\begin{tabular}{c|c|c|c|c||c|c|c|c|c}
\hline
\multirow{2}{*}{ID} & \multirow{2}{*}{RA} & \multirow{2}{*}{Dec} & Number& Number~~ & \multirow{2}{*}{ID} & \multirow{2}{*}{RA} & \multirow{2}{*}{Dec} & Number& Number~~ \\
&&  & of galaxies & of clusters&&&  & of galaxies & of clusters \\
\hline
1 &   $2^{\rm h}44^{\rm m}00^{\rm s}$   &   +41\degr00\arcmin00\arcsec   & 13221 & 73 & 6 &   $3^{\rm h}00^{\rm m}00^{\rm s}$   &   +43\degr00\arcmin00\arcsec    & ~~9844  & 37  \\
2 &   $2^{\rm h}52^{\rm m}00^{\rm s}$   &   +41\degr00\arcmin00\arcsec   & 10965 & 59 & 7 &   $2^{\rm h}44^{\rm m}00^{\rm s}$   &   +45\degr00\arcmin00\arcsec    & 14085 & 105 \\
3 &   $3^{\rm h}00^{\rm m}00^{\rm s}$   &   +41\degr00\arcmin00\arcsec   & 10751 & 37 & 8 &   $2^{\rm h}52^{\rm m}00^{\rm s}$   &   +45\degr00\arcmin00\arcsec    & 15283 & 120 \\
4 &   $2^{\rm h}44^{\rm m}00^{\rm s}$   &   +43\degr00\arcmin00\arcsec   & 12587 & 75 & 9 &   $3^{\rm h}00^{\rm m}00^{\rm s}$   &   +45\degr00\arcmin00\arcsec    & 11780 & 73  \\
5 &   $2^{\rm h}52^{\rm m}00^{\rm s}$   &   +43\degr00\arcmin00\arcsec   & 12430 & 85 &10 &   $3^{\rm h}00^{\rm m}00^{\rm s}$   &   +47\degr00\arcmin00\arcsec    & 12374 & 83  \\
\hline
\end{tabular}
\end{table*}

The first galaxy cluster catalogs were compiled by Abell in
1958~\cite{Abell1958:Grohovskaya_n_en} and Zwicky over the period
of 1961 to 1968~\cite{Zwicky1961:Grohovskaya_n_en} by visually
inspection the Palomar survey, whereas in the 1980's fully
automated methods for isolating cluster structures became
available
(e.g.,~\cite{Shectman1985:Grohovskaya_n_en,Dodd1986:Grohovskaya_n_en}).

In our work we used methods involving Voronoi tesselation and
adaptive kernel to analyze the large scale galaxy distribution and
determine candidates for cluster structures.

\begin{figure}[bpt!!!]
\includegraphics[width=1.0\columnwidth]{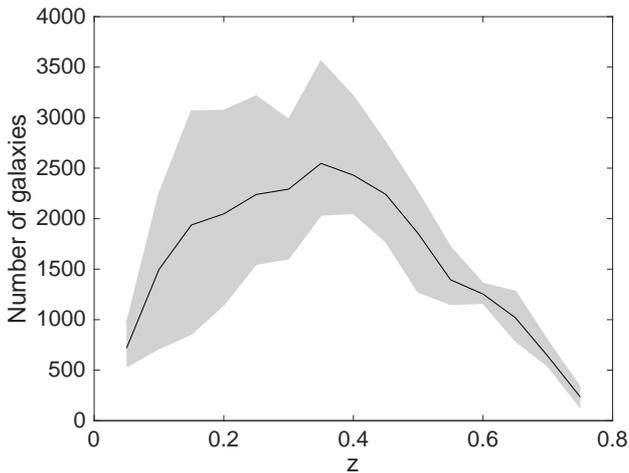}
\caption{Dependence of the number of galaxies on redshift. For $z
\sim 0.3$--$0.4$ a sharp increase is evident in the number of
galaxies in the layer, due to a fast increase in the light cone
volume. The line corresponds to the average number of galaxies in
the layer for 10~galaxy data sets from the MICE simulation, and
the shaded regions shows the data scatter. } \label{fig_1:Grohovskaya_n_en}
\end{figure}

\begin{figure*}[bpt!!!]
\hspace*{50pt}
\includegraphics[scale=0.5]{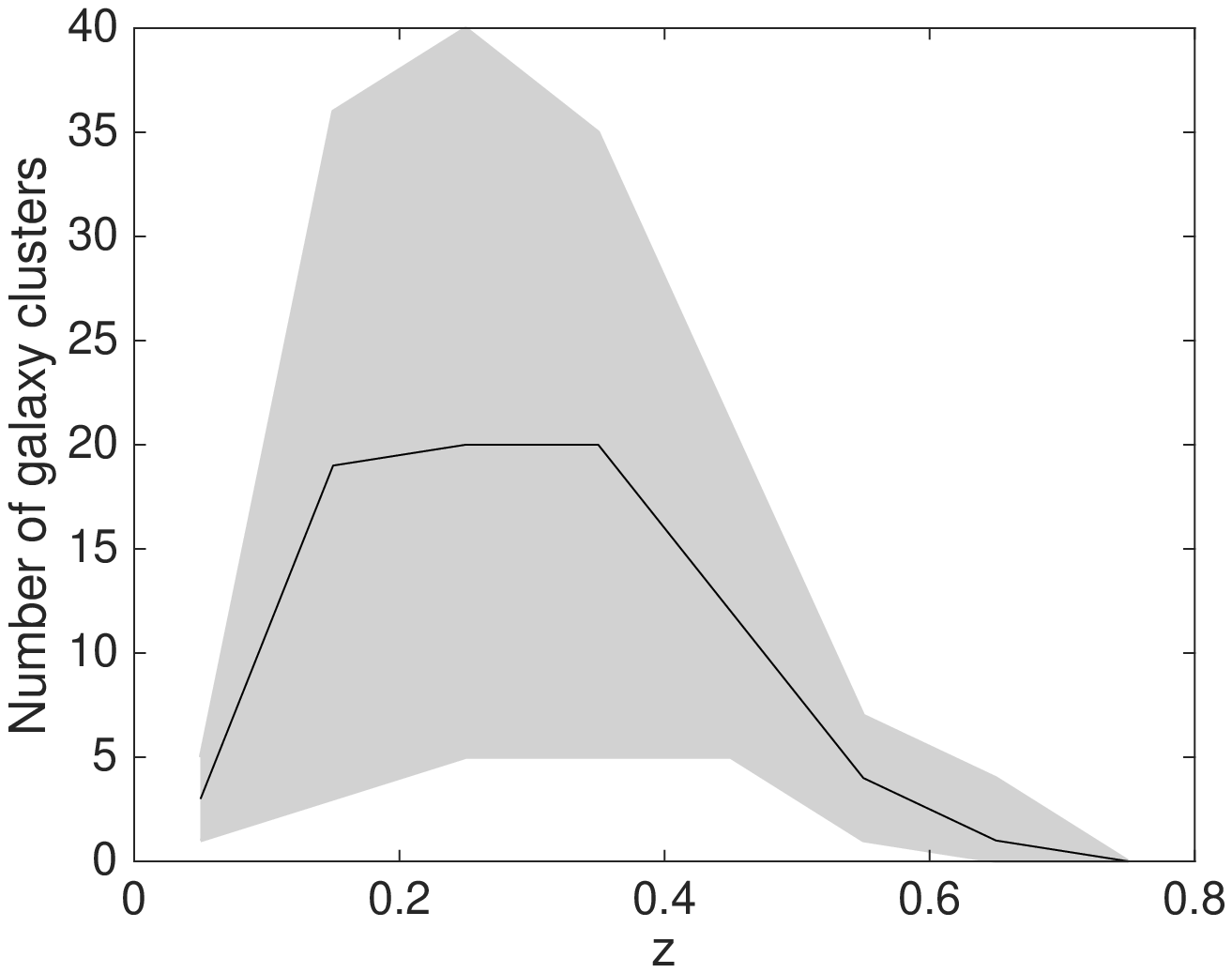} 
\includegraphics[scale=0.5]{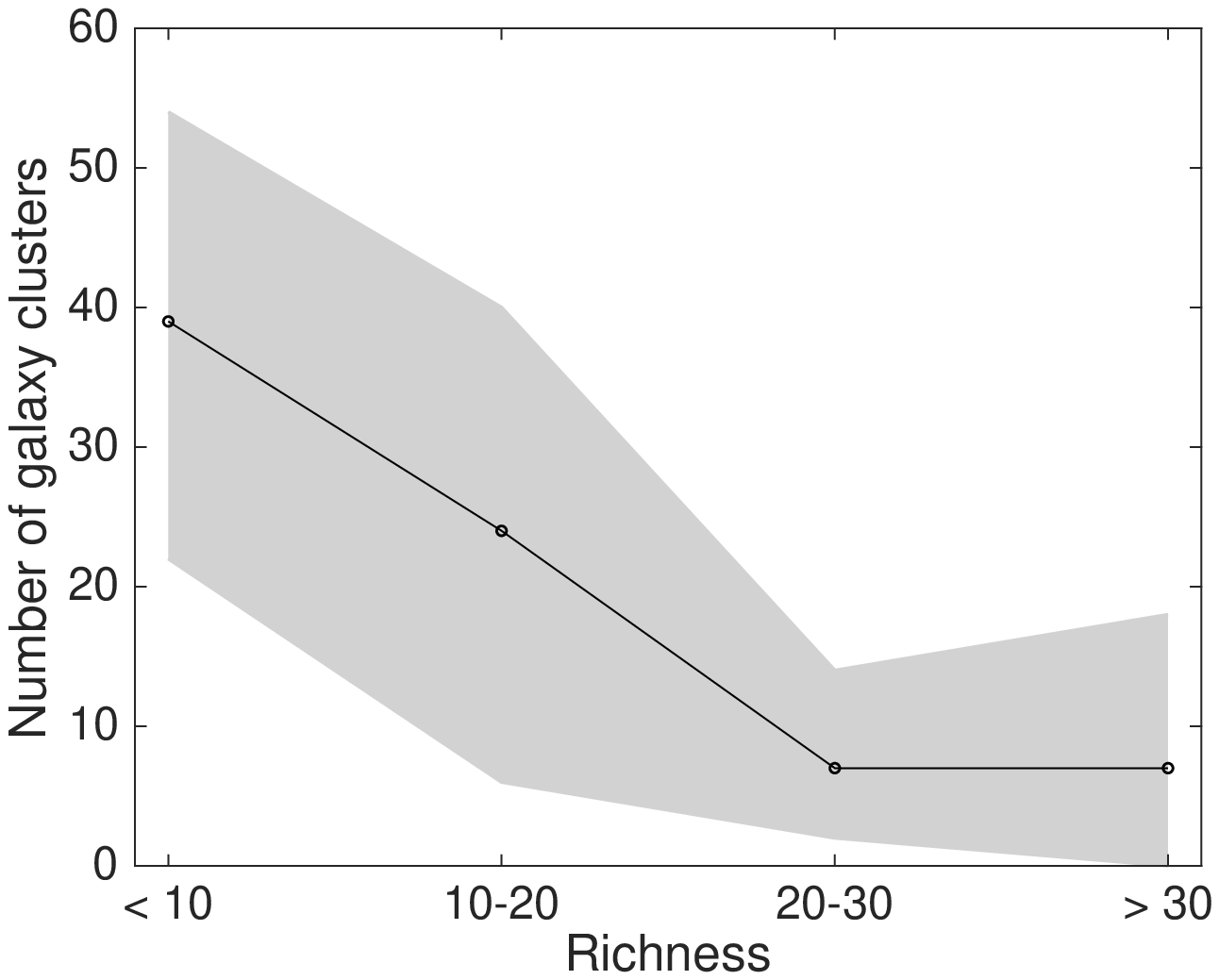} 
\caption{Relations between the number of galaxy clusters and
redshift (a) and cluster richness class (b). The continuous lines
show the average number of galaxies for 10 galaxy data sets from
the MICE simulation, the shaded region indicates the data scatter
which shows that the sets for algorithm testing have various
distributions of galaxy clusters by redshift. }
\label{fig_2:Grohovskaya_n_en}
\end{figure*}

We tested the developed methods of determining the composition of
the large scale galaxy distribution on 10 MICE
collaboration~\cite{Carretero2017:Grohovskaya_n_en} data sets from
lightcone $N$~body simulation. All galaxy samples are restricted
by an area of 2 square degrees, a limiting magnitude of $R_{\rm
AB} = 23^{\rm m}$ and redshift from~$0$ to~$0.8$, which
corresponds to the limitations of the observed data which will be
analyzed in the future. The coordinates on the sample field
centers, as well as the number of galaxies and their clusters for
each of the 10~data sets are presented in the Table.
We also constructed plots of the distribution of the number of
galaxies as a function of
redshift~(Fig.~\ref{fig_1:Grohovskaya_n_en}) and the distribution
of the number of galaxy clusters as a function of redshift and
cluster richness~(Fig.~\ref{fig_2:Grohovskaya_n_en}). The plots
demonstrate a rapid increase in the number of galaxies and
clusters at \mbox{$z \sim 0.3$--$0.4$} and a decrease at \mbox{$z
\sim 0.7$--$0.8$}, determined by the threshold magnitude of the
samples. The majority of the galaxies which are related to large
scale structures are collected into groups in the studied samples,
and only a few are related to clusters.

The data of the MICECAT mock catalog are convenient for testing
algorithms of large scale structure selection, since they contain
information on the dark matter halo which a galaxy belongs to.
Correspondingly, galaxies with the same halo identifier belong to
the same cluster and are gravitationally bound.

Since the methods of analysis were developed for subsequent work
with the data obtained with the \mbox{1-m}~Schmidt telescope of
BAO NAS, the accuracy of photometric redshift determination for
which is $0.01$, candidates to various structures were determined
for narrow layers of the large scale structure with a step of
\mbox{$\Delta z = 0.02 (1 + z)$} for all methods; we also added to
this interval 25\% of its value on each side, in order to avoid
losses in determining galaxy clusters that are positioned on the
boundary between the two layers.

\subsection{Algorithm of Determining the Density Contrast with Adaptive Kernel}

\label{ss_2.1:Grohovskaya_n_en}
\begin{figure*}[bpt!!!]
\hspace*{30pt}
\includegraphics[width=0.85\columnwidth]{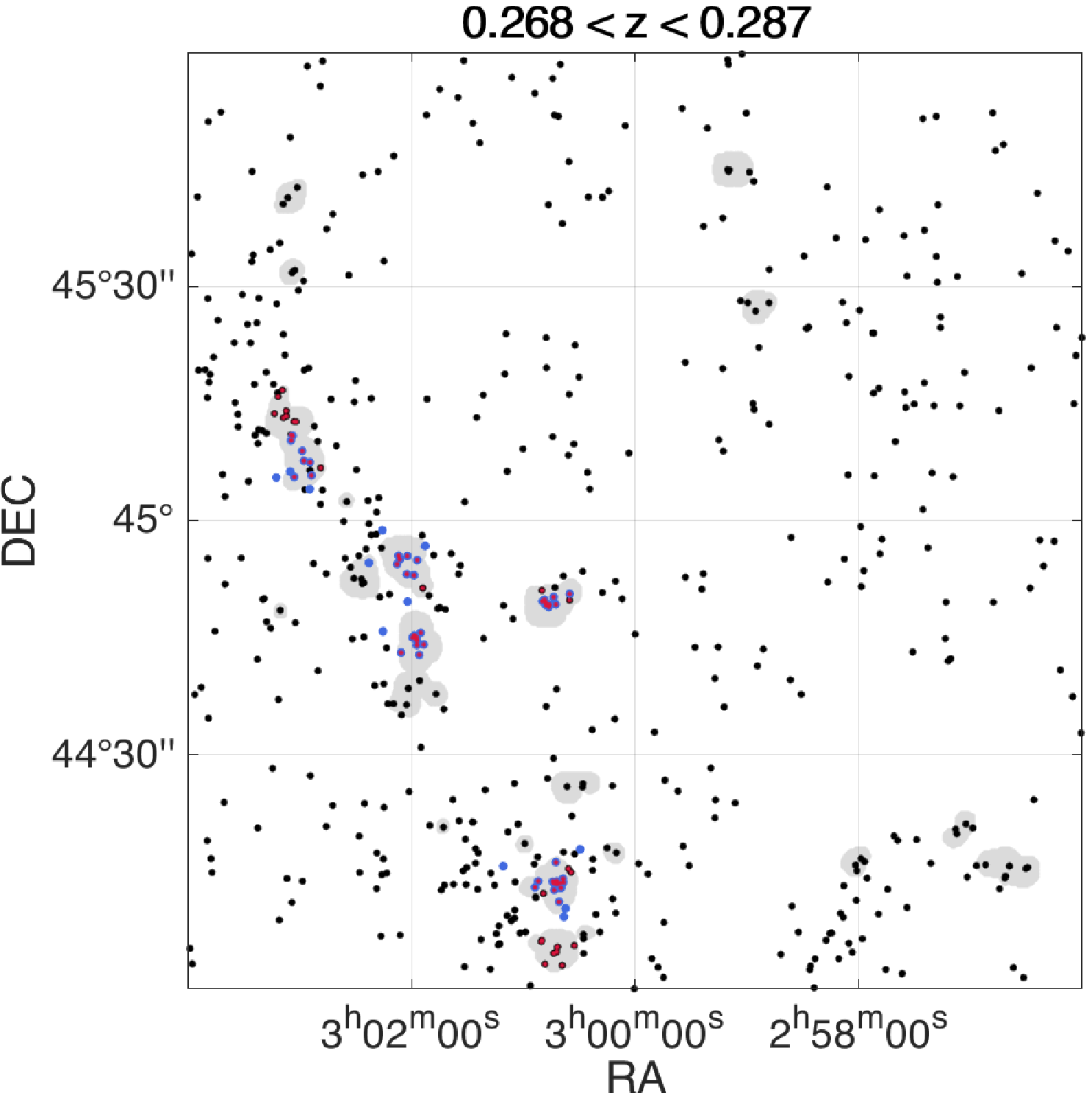}
\includegraphics[width=0.85\columnwidth]{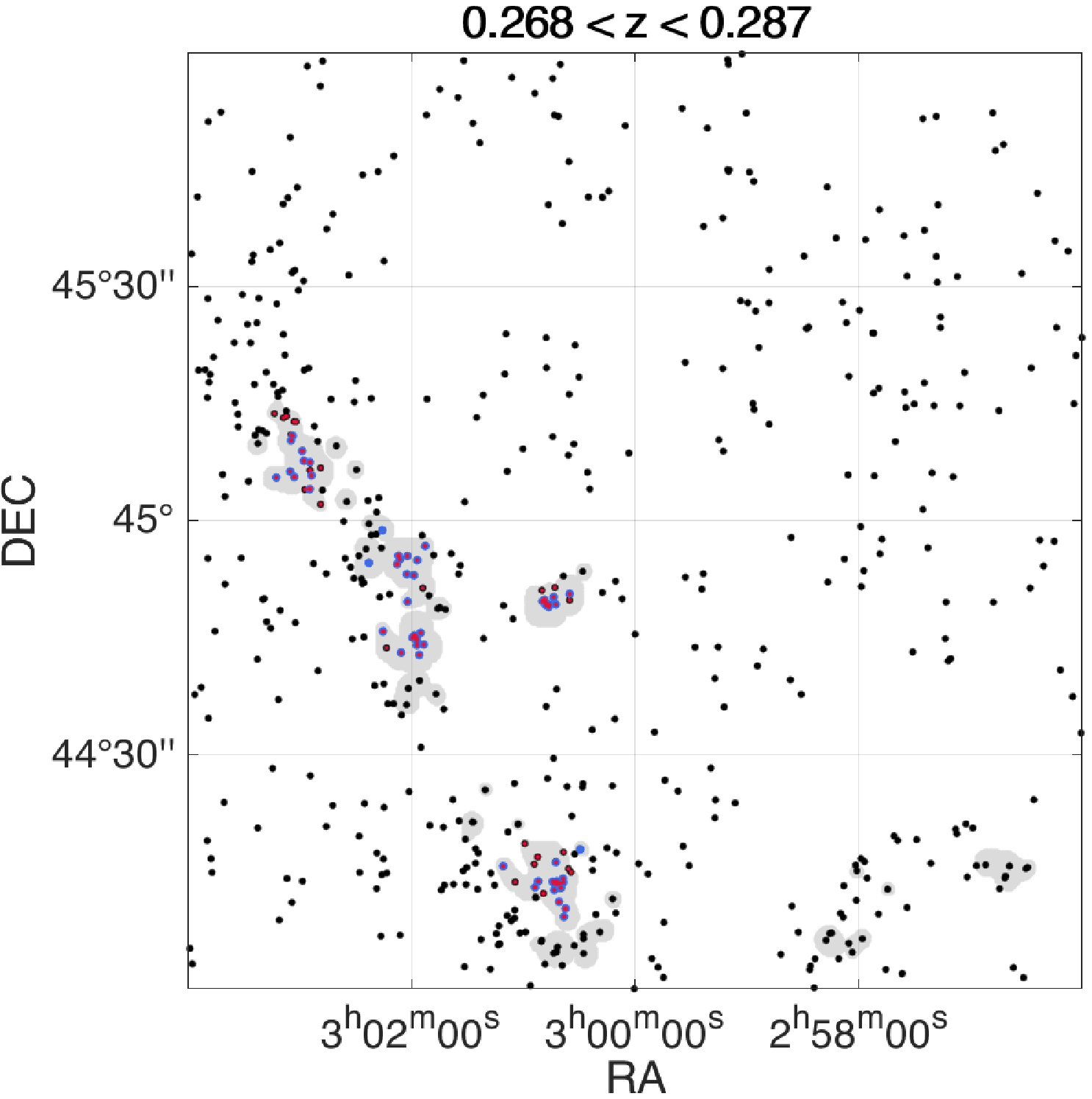}
\hspace*{85pt}
\includegraphics[width=0.85\columnwidth]{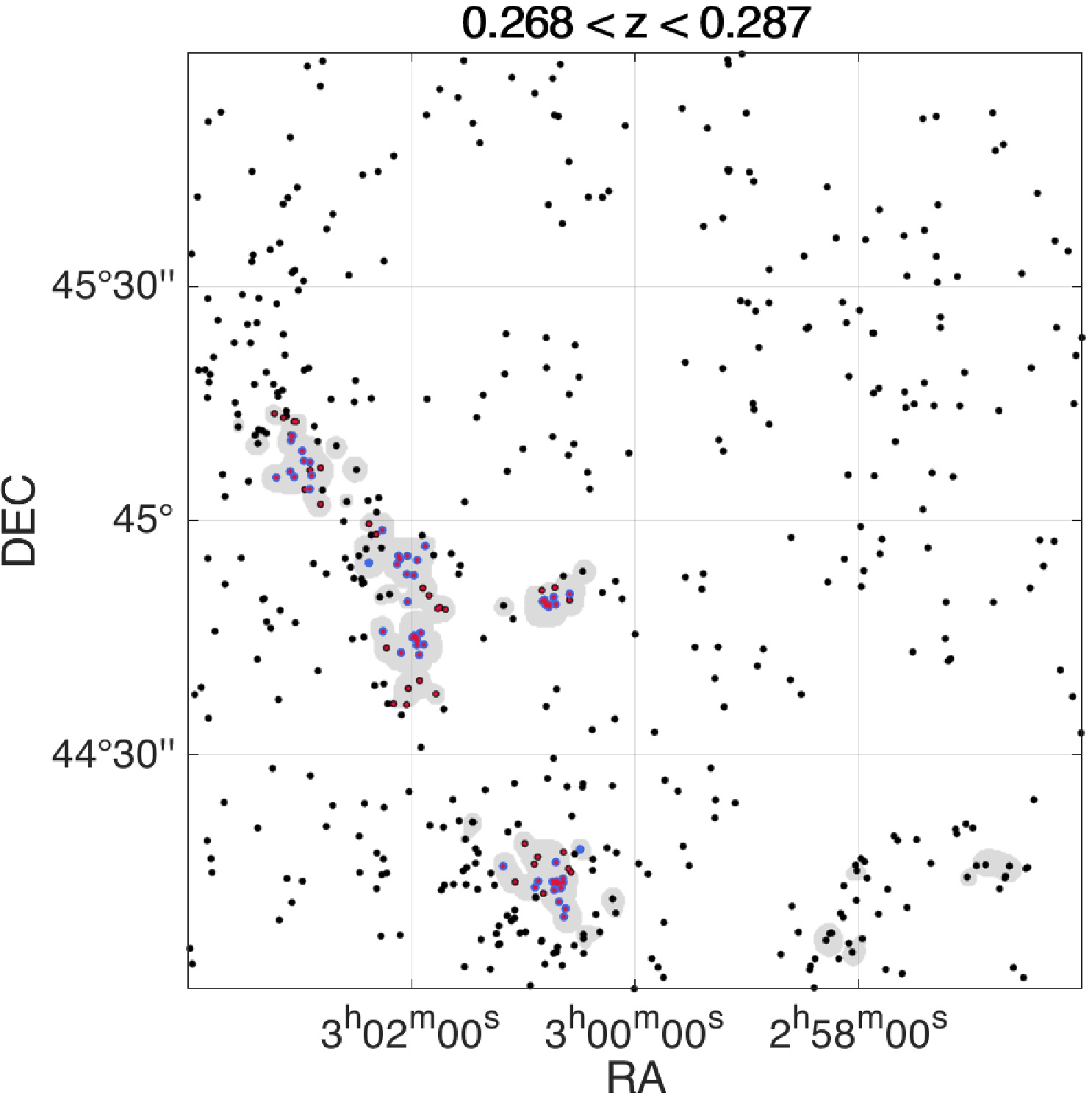}
\includegraphics[width=0.85\columnwidth]{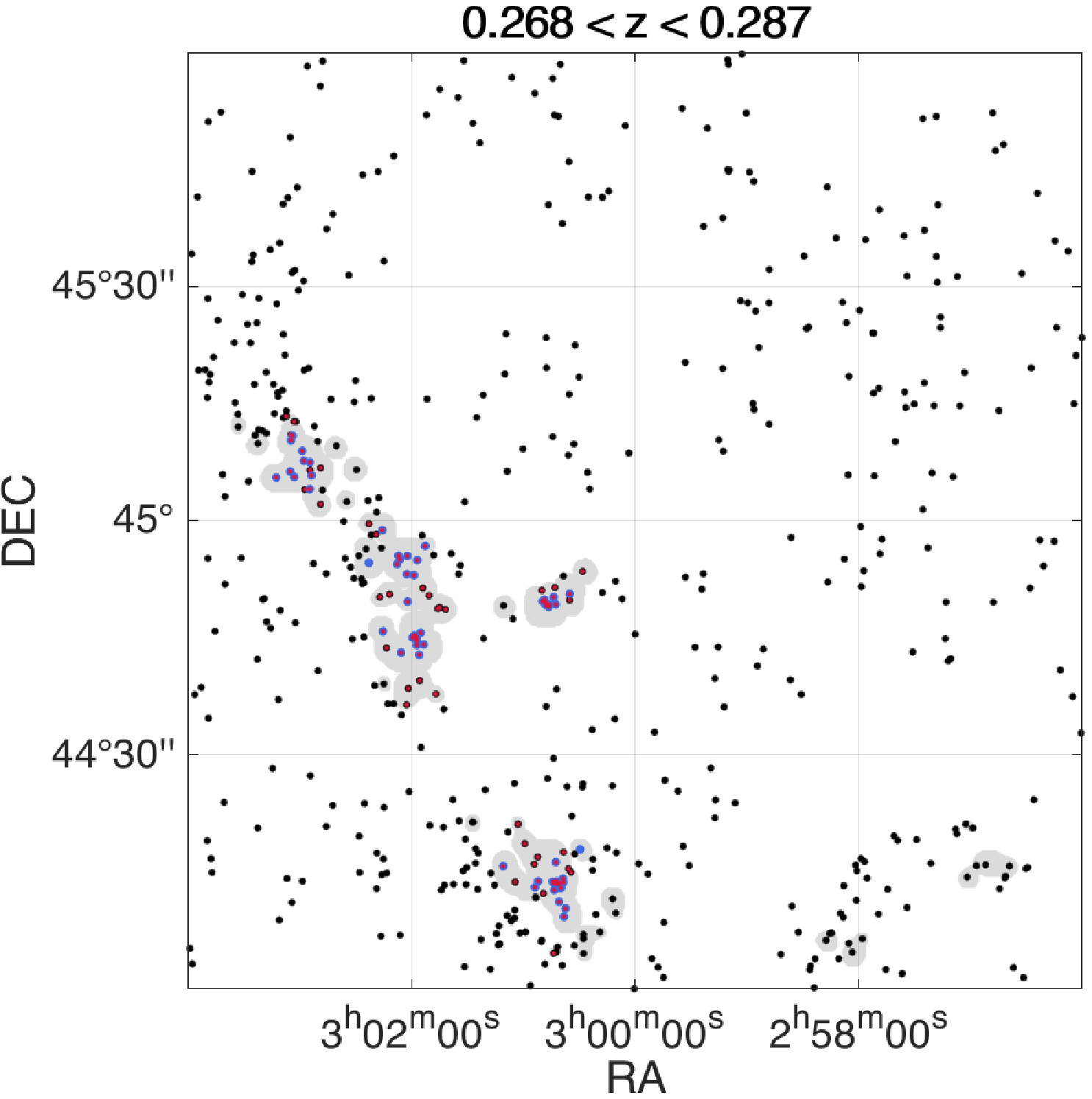}
\hspace*{85pt}
\includegraphics[width=0.85\columnwidth]{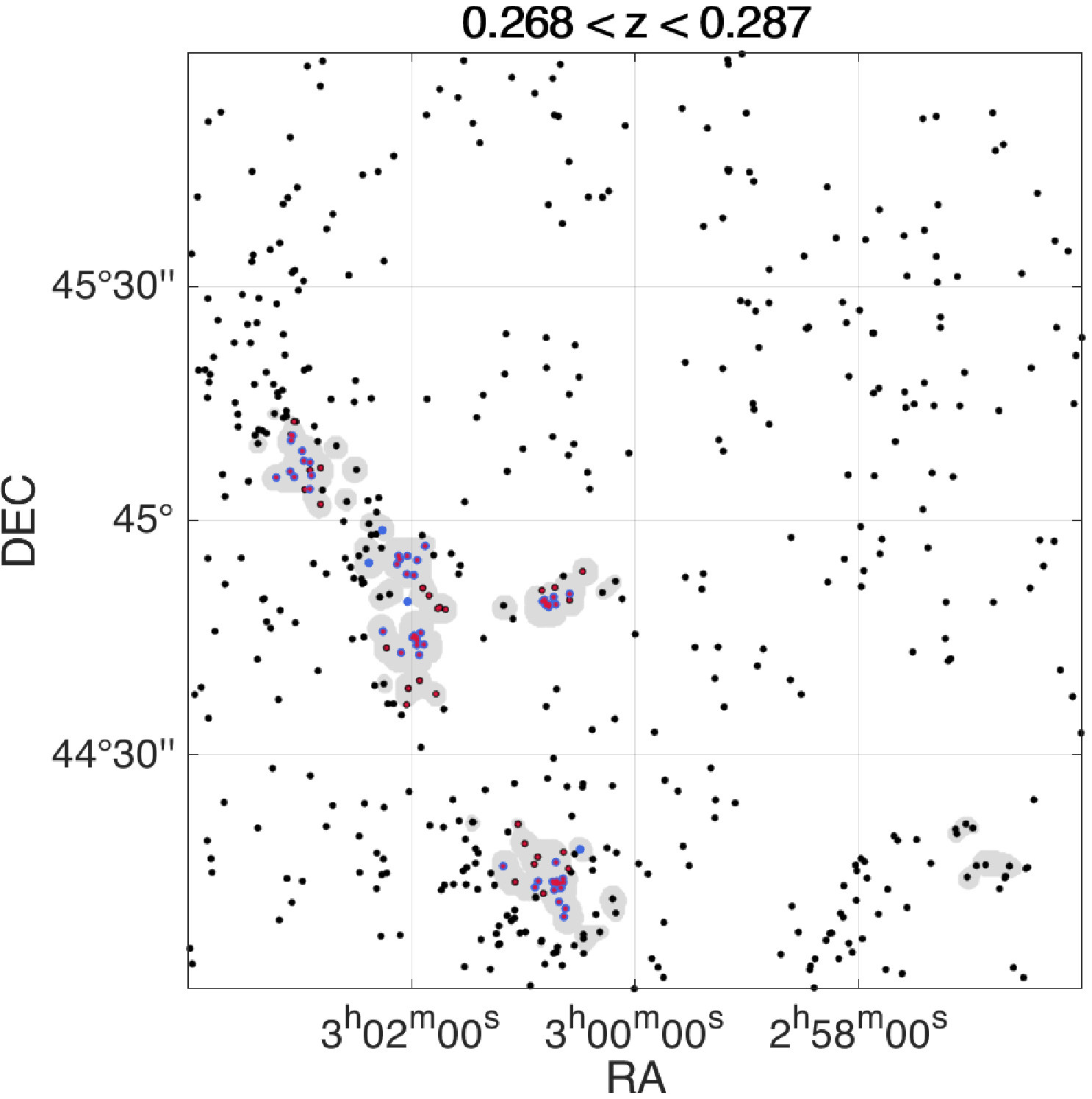}
\includegraphics[width=0.85\columnwidth]{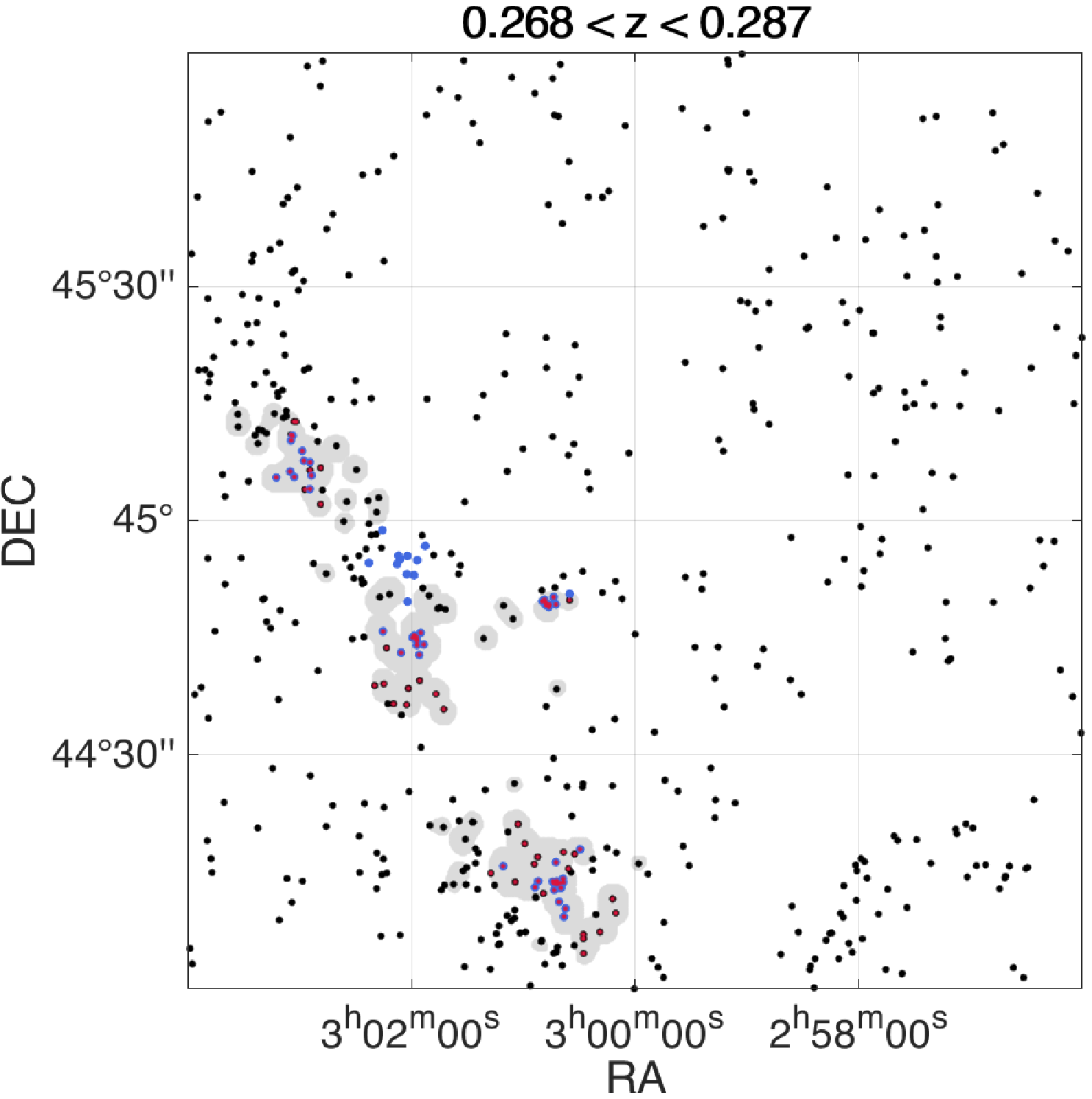}
\caption{Density contrast maps for two dimensional layers of the
three dimensional large scale distribution of galaxies, obtained
by using adaptive kernel up to the 2-nd (a), 5-th (b), 7-th (c),
8-th (d), \mbox{10-th~(e)} and 25-th neighbor~(f). The gray color
marks regions where the density contrast exceeds $2$. The blue
dots show galaxies that belong to the same dark matter halo in the
model catalog; red shows galaxies located in software-determined
clusters. } 
\label{fig_3:Grohovskaya_n_en}
\end{figure*}

\begin{figure}[]
\includegraphics[width=\columnwidth]{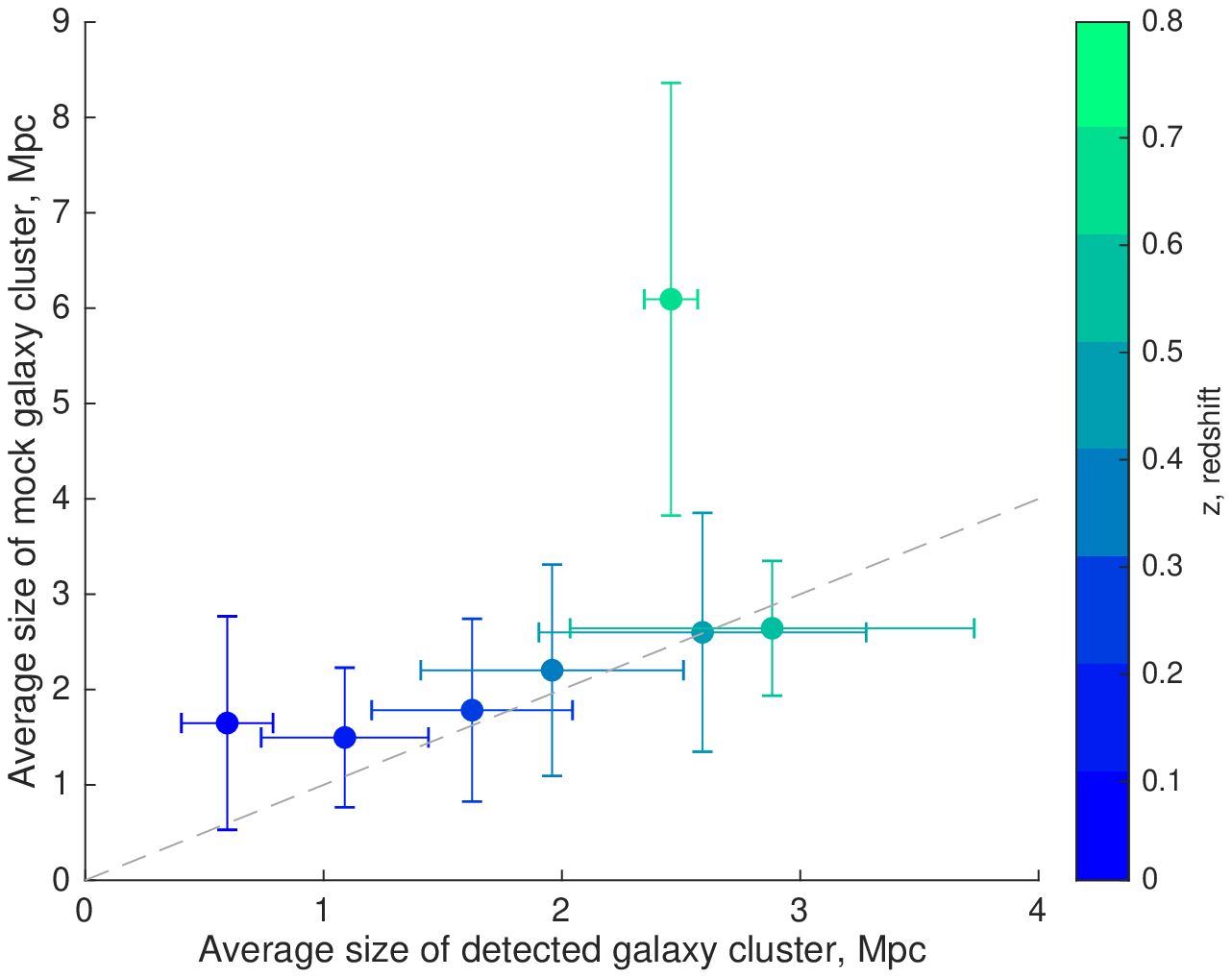}
\caption{The size of the galaxy clusters detected by a filtering
algorithm with adaptive kernel, as a function of cluster size in
the model catalog. The color scale corresponds to the redshift,
and the bars show the standard deviations from the average size of
the detected and model cluster at each $z$. The values are given
for 10 parameter sets from the MICE simulation. } \label{fig_4:Grohovskaya_n_en}
\end{figure}

The galaxy distribution density was determined in the neighborhood
of each considered galaxy as:
\begin{equation}
\delta_i=\dfrac{s}{\frac{4}{3} \pi R^3}.
\end{equation}
The size of the area $R$, or the adaptive kernel which was used to
smooth the galaxy density, was determined by the distance from the
galaxy to the $s$-th nearest neighbor as a three dimensional
distance between the $i$-th galaxy and its $j$-th
neighbor~\cite{Kibble2004:Grohovskaya_n_en}:
\begin{equation}
R=\sqrt{(X_i-X_j)^2+(Y_i-Y_j)^2+(Z_i-Z_j)^2},
\end{equation}

\noindent where the Cartesian coordinates $X, Y, Z$ are computed
by the formulas of conversion from spherical to Cartesian
coordinate systems:
\begin{eqnarray}
X_i=R_{i}  r_i  \sin{\Theta_i} \cos{\phi_i},  \nonumber\\
Y_i=R_{i}  r_i  \sin{\Theta_i}  \sin{\phi_i},\\
Z_i=R_{i}  r_i  \cos{\Theta_i}.\nonumber
\end{eqnarray}
The angles $\Theta_i=({\dfrac \pi 2}-{\rm Dec})$, $\phi_i={\rm
RA}$ (in radians); \mbox{$R_{i}= (1+z_i)^{-1}$}---scaling factor,
$z_i$---redshift of the $i$-th galaxy; $r_i$---the comoving
distance of the $i$-th galaxy is calculated according to the
formula from~\cite{Peacock1999:Grohovskaya_n_en}:
\begin{equation}
r_i= \dfrac {c}{H_0} \int \limits_0^{z_i }{(\Omega_M \cdot
(1+z_i)^3+1-\Omega_M)^{-0.5}}dz.
\end{equation}

Clusters and groups of galaxies are defined as peaks on density
contrast maps of the distribution of galaxies. To mark the
positions of the density peaks we determine the average density in
a slice
\begin{equation}
\overline \delta = {\frac 1 n}  \sum_{i=1}^n \delta_i.
\end{equation}
Here $n$ is the number of galaxies in each slice. We then compute
the value showing the density contrast in each point
\mbox{$\sigma_i+1=\dfrac{(\delta_i-\overline \delta)}{\overline
\delta} +1$} and interpolate the density contrast values over the
entire field.

In this work we investigated the use of adaptive apertures up to the 2nd, 5th, 7th, 8th, 10th and 
20th neighbor~(Fig.~\ref{fig_3:Grohovskaya_n_en}). Statistical
estimates (see Section~\ref{ss_2.4:Grohovskaya_n_en}) show that
increasing the size of the adaptive kernel leads to the blurring
of peaks on density contrast maps of the galaxy distribution and
to the under-estimation of the large scale structures. We selected
the distance to the 8th neighbor as a operating aperture of the
method, which best allows us to isolate various structures.

For the operating aperture we also estimated the average sizes of
the detected clusters depending on the average sizes of the
clusters in the mock catalog in redshift slices
(Fig.~\ref{fig_4:Grohovskaya_n_en}) corresponding to the current
notions about the sizes of galaxy clusters from~2 to 5~Mpc. Only
two clusters fall within the range of \mbox{$0.6 \leq z \leq 0.7$}
among the 10 MICE catalog samples, which explains the large data
scatter in this point. For $0.7 \leq z \leq 0.8$ the galaxy
clusters are absent in the catalog due to the $R_{\rm AB}$
limitations and the number of cluster members (more than eight);
there are also no detected clusters there.

\subsection{Voronoi Tesselation Algorithm for Determining the Density Contrast}
\label{ss_2.2:Grohovskaya_n_en}

A Voronoi tesselation is defined as such a geometric devision of a
plane into a polygon which possesses the following property: for
any center $p_i$ of a system of points there is a region of space
(polygon or Voronoi region) where all points are closer to the
given center than to any other system
center~\cite{Ramella2001:Grohovskaya_n_en}. In this work, we use
the considered galaxies as Voronoi region peaks. The procedure of
dividing a two dimensional projection of a layer of the three
dimensional large scale galaxy distribution was carried out with
the procedures {\tt `voronoi'} and {\tt `voronoi'} in MatLab
environment.

The inverse area of a Voronoi region is the numerical density
corresponding to each galaxy---a polygon vertex. Groups and
clusters of galaxies are determined in a way similar to the
previous method, by the peaks on the density contrast map for the
galaxy distribution field. The density contrast is defined as
$\sigma_i=\dfrac{(\delta_i-\overline \delta)}{\overline \delta}$,
and the average slice density:
\begin{equation}
\overline \delta = {\dfrac{1}{n}}  \sum_{i=1}^n A_i,
\end{equation}
\noindent where $A_i$ is the area of the Voronoi polygon around
the $i$-th galaxy, and $n$ is the number of Voronoi cells in the
considered layer~\cite{Sochting2012:Grohovskaya_n_en}. When
computing the average density we do not take into account the
boundary points for which the Voronoi cells tend to infinity.

\begin{figure}[bpt!!!]
\includegraphics[width=0.9\linewidth]{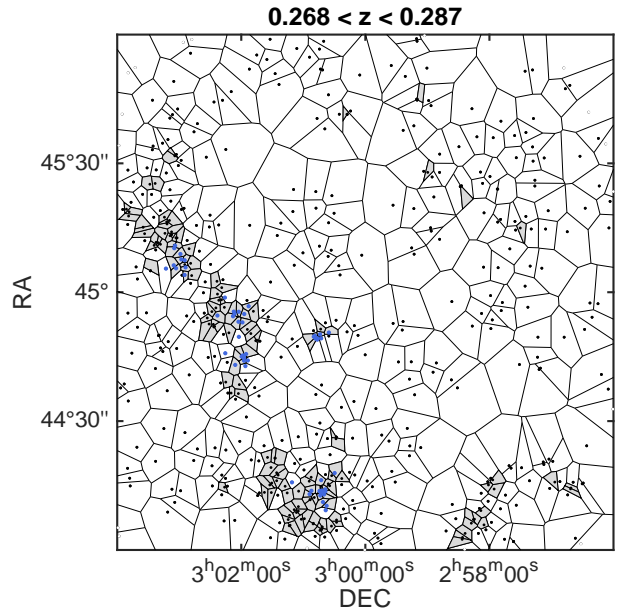}
\caption{Voronoi division for a two dimensional thin layer of the large scale distribution of galaxies. The grey color shows cells whose density exceeds the average density of the layer more than twice. Galaxies that belong to the same dark matter halo in the model catalog are marked by blue. 
} \label{fig_5:Grohovskaya_n_en}
\end{figure}

\subsection{ Determining Groups and Clusters of Galaxies}
\label{ss_2.3:Grohovskaya_n_en} We selected regions with a density
contrast of \linebreak$\sigma_i+1 \geq 2$ as galaxy group and
cluster candidates. For the algorithm using adaptive kernel a
galaxy group or cluster candidate must contain more than nine
galaxy members inside the region with the density contrast greater
than $2$, and for the Voronoi algorithm it must fulfill the
condition of at least eight other cells with a density contrast
higher than $2$ located near the cell (see
Section~\ref{ss_2.4:Grohovskaya_n_en}). An example of dividing a
narrow slice of the large scale redshift distribution with marked
detected clusters and clusters from the catalog is presented in
Fig.~\ref{fig_5:Grohovskaya_n_en}.

\begin{figure*}[bpt!!!]
\hspace*{45pt}
\includegraphics[scale=0.27]{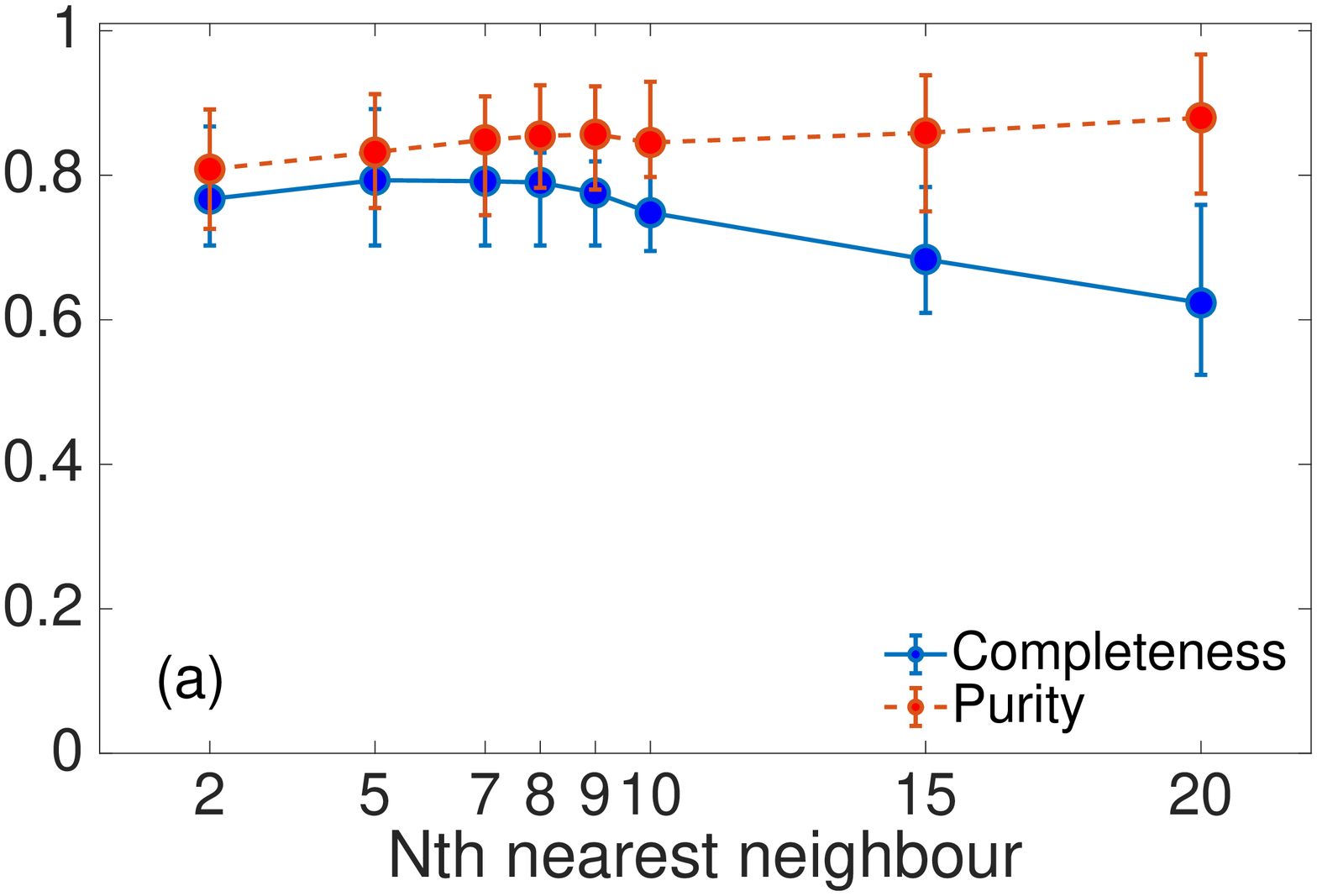}
\includegraphics[scale=0.27]{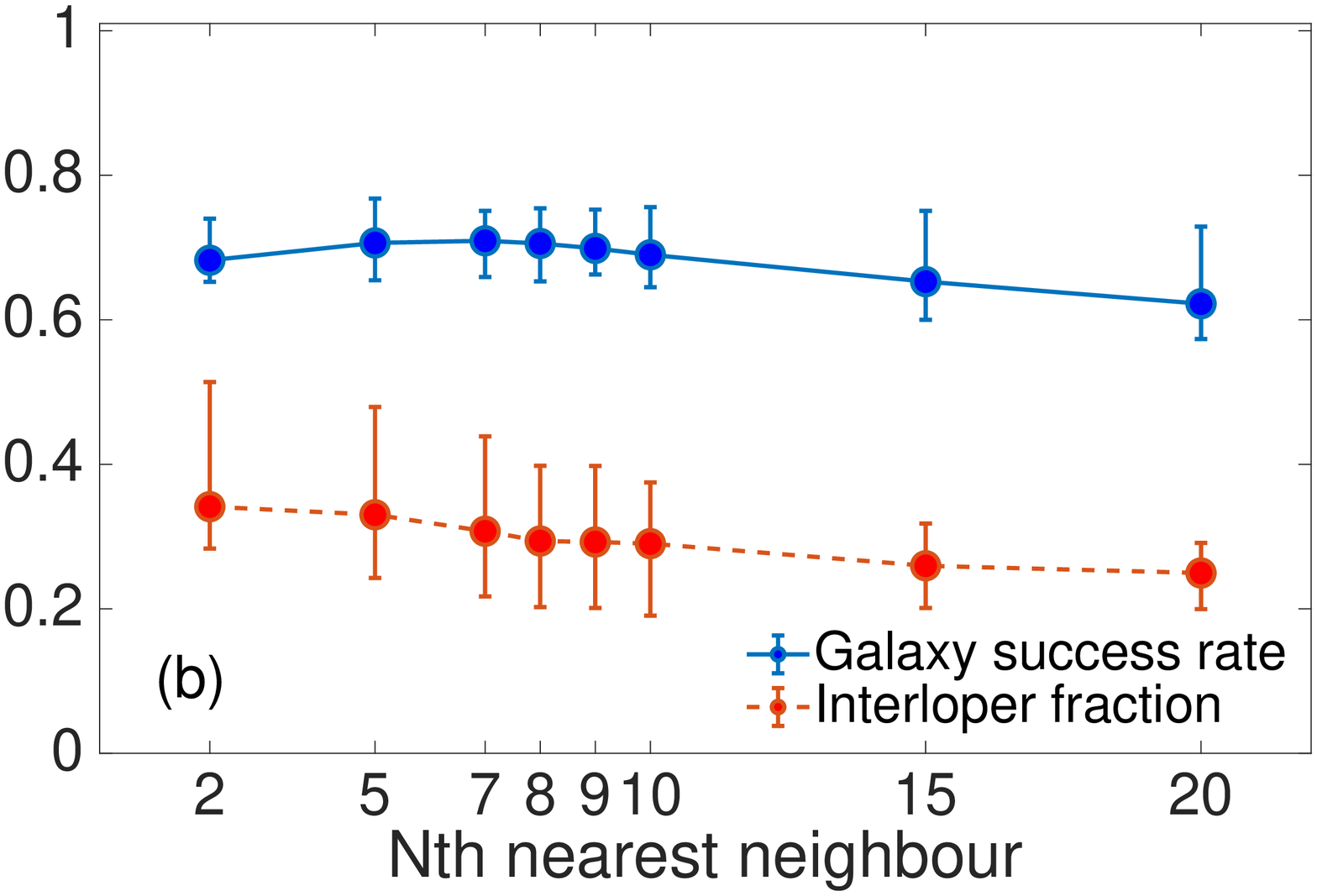}
\\
\hspace*{45pt}
\includegraphics[scale=0.27]{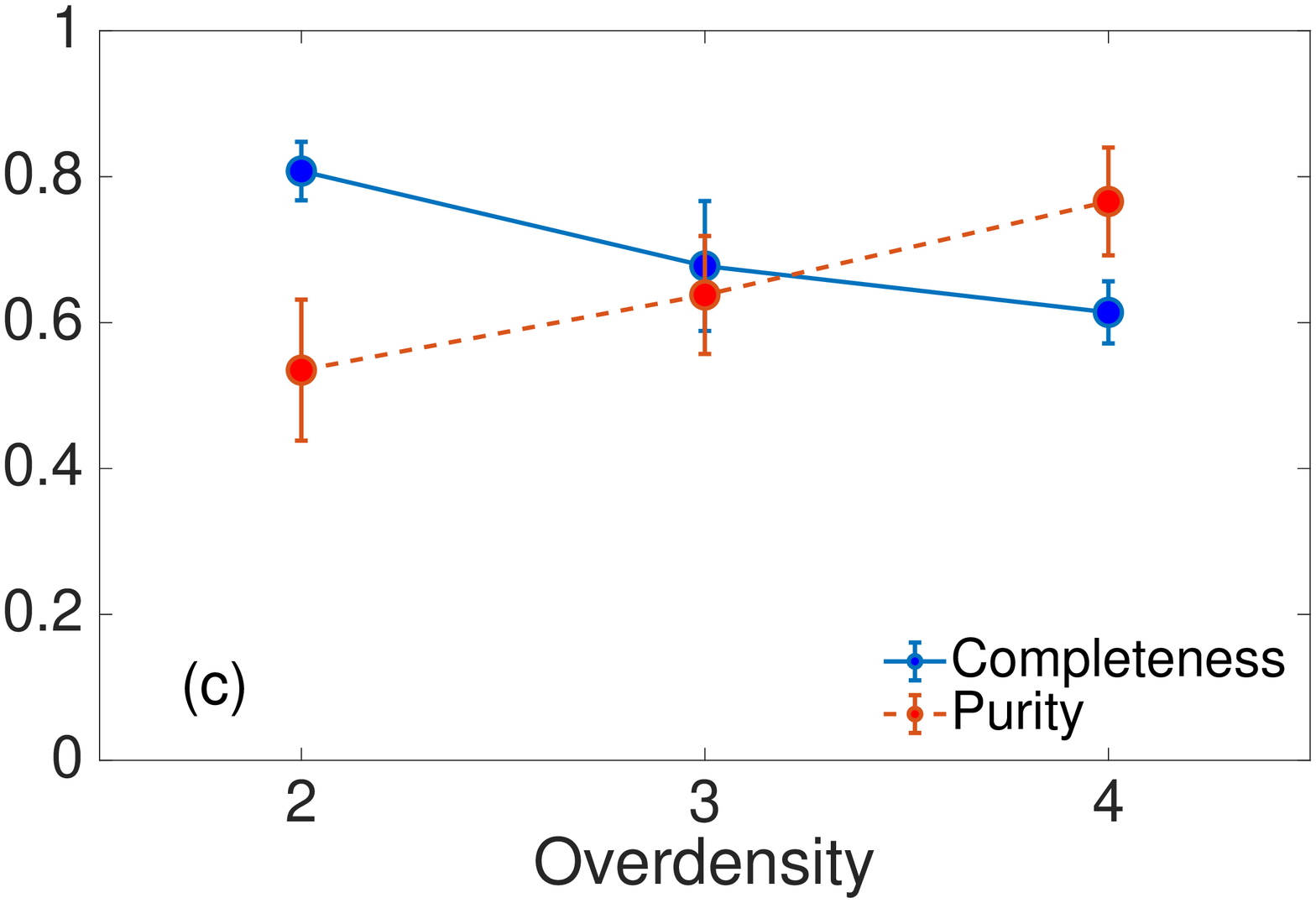}
\includegraphics[scale=0.27]{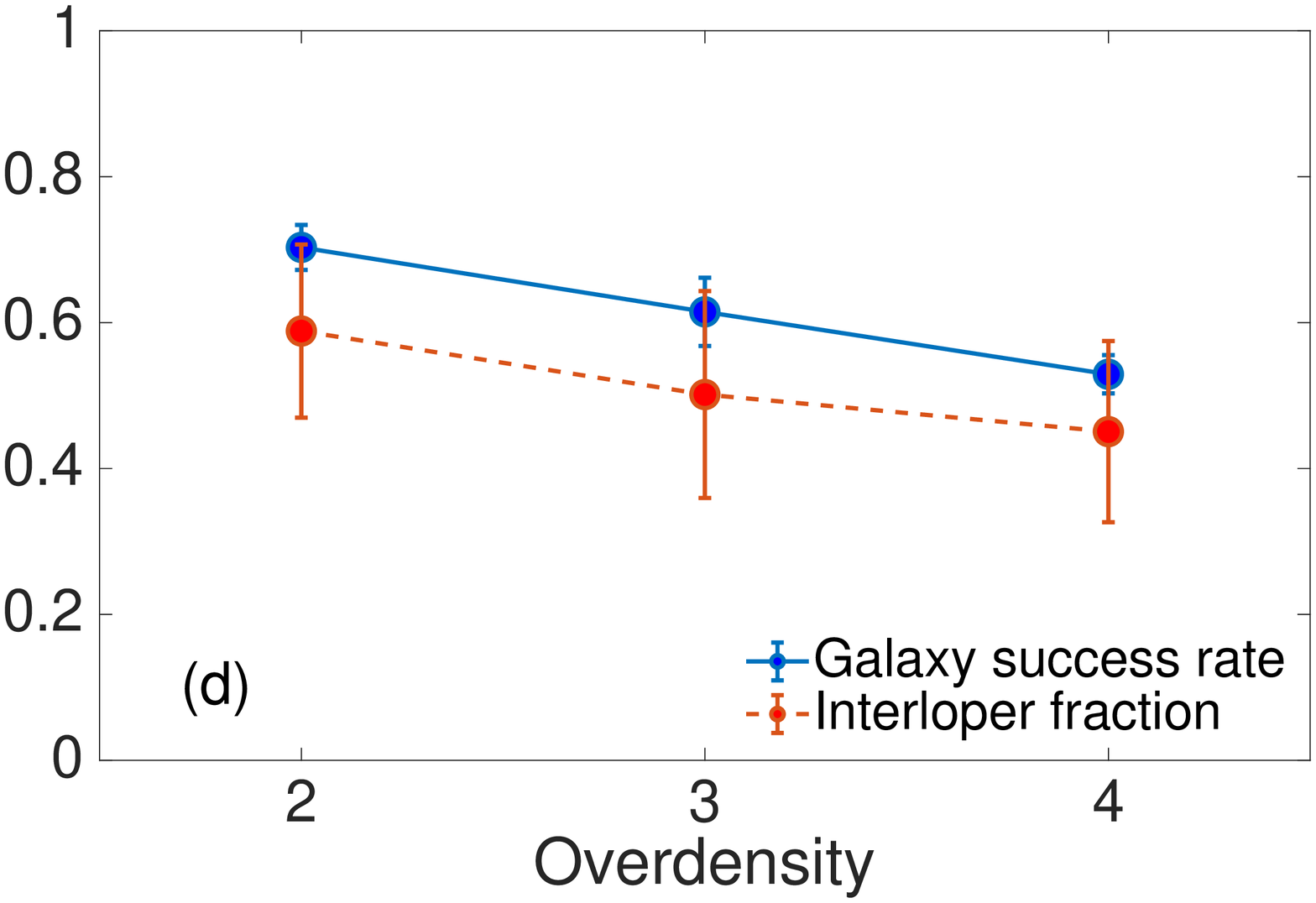}
\caption{Statistical parameters of a sample of groups, clusters
and their galaxies, obtained by software algorithms. The upper
figures are given for the filtering algorithm with adaptive
kernel, and the lower figures represent the Voronoi tesselation
algorithm. The left hand side figures show the completeness (blue
solid line) and purity (red dashed line) of the sample, and on the
right we give the percentage of successful identifications of
galaxies from the model catalog clusters (blue solid line) and the
percentage of galaxies identified by the software as cluster
galaxies, but which are listed as field galaxies in the model
catalog (red dashed line). The bars show the root-mean-square
deviations for 10 sets of parameters from the MICE simulation.}
  \label{fig_6:Grohovskaya_n_en}
\end{figure*}

On the whole, both methods of analyzing the large scale
distribution of galaxies give compatible results. However, the
Voronoi tesselation method, despite the high fraction of detected
cluster member galaxies and identified clusters in the catalogs in
general, shows a much smaller sample purity both in terms of
clusters and galaxies (see Fig.~\ref{fig_6:Grohovskaya_n_en}). We
can thus conclude that the Voronoi tesselation method may be
useful as an additional tool for identifying clusters, whereas the
adaptive kernel algorithm should be used as the main instrument.

\subsection{ Basic Statistical Calculations}
\label{ss_2.4:Grohovskaya_n_en} Since the MICE galaxy mock catalog
has a parameter that determines the cluster membership of a
galaxy, we can estimate the quality of the developed algorithms
using statistical methods. We used the statistical estimation
criteria
from~\cite{Gerke2005:Grohovskaya_n_en,Knobel2010:Grohovskaya_n_en}.
According to these papers, group $i$ is considered to correspond
to group $j$ if group $j$ contains more than $f\%$ of group $i$
galaxies. From this we can determine that the correspondence can
be one sided when group $i$ corresponds to group $j$ and the
reverse is false, and mutual when group $i$ corresponds to group
$j$ and vice versa. In this work we assume that group $i$
corresponds to group $j$ if group $j$ contains more than $50 \%$
of galaxies belonging to $i$.

The terms of sample completeness and purity can also be defined
for the one sided and two sided correspondence. Let us denote by
$N_{\rm real}^{\rm gr}$ the number of galaxy groups and clusters
in the model catalog, and let $N_{\rm rec}^{\rm gr}$ represent the
number of groups and galaxies determined by the program
algorithms. Then the  ``one-sided sample completeness'' of the
clusters corresponding to the dark matter halo of the model
catalog clusters is computed as
\begin{equation}
c_1 =\dfrac{N_{\rm real}^{\rm gr}\rightarrow N_{\rm rec}^{\rm
gr}}{N_{\rm real}^{\rm gr}},
\end{equation}
where $N_{\rm real}^{\rm gr}\rightarrow N_{\rm rec}^{\rm gr}$ is
the number of matches between the real groups from the catalog and
the groups determined automatically.

The ``two-sided sample completeness'':
\begin{equation}
c_2 =\dfrac{N_{\rm real}^{\rm gr}\leftrightarrow N_{\rm rec}^{\rm
gr}}{N_{\rm real}^{\rm gr}},
\end{equation}
here $N_{\rm real}^{\rm gr}\leftrightarrow N_{\rm rec}^{\rm gr}$
is the number of matches of real groups from the catalog and
groups determined automatically and vice versa. The purity of the
sample can be determined in a similar way:
\begin{equation}
p_1 =\dfrac{N_{\rm rec}^{\rm gr}\rightarrow N_{\rm real}^{\rm
gr}}{N_{\rm rec}^{\rm gr}},
\end{equation}
\begin{equation}
p_2 =\dfrac{N_{\rm rec}^{\rm gr}\leftrightarrow N_{\rm real}^{\rm
gr}}{N_{\rm rec}^{\rm gr}}
\end{equation}
The four quantities can take only values between $0$ and $1$. A
close to $1$ value of parameter $c_1$ shows that the number of
undefined groups in the catalogs is small. For parameter $p_1$
such a value shows a small fraction of falsely detected clusters.
The $c_1$ to $c_2$ and $p_1$ to $p_2$ ratios close to zero show an
overestimation (one automatically detected group corresponds to
several groups in the catalog) or fragmentation (one catalog group
corresponds to several automatically identified groups) of the
groups.

The statistical parameters $c_1$, $c_2$, $p_1$ and $p_2$
characterize the work of program algorithms for groups and
clusters of galaxies. In order to estimate it for the galaxies in
groups, let us denote the sample of galaxies in groups and
clusters in the model catalog as $S_{\rm real}^{\rm gal}$, and the
sample of galaxies in the automatically identified clusters as
$S_{\rm rec}^{\rm gal}$. We then introduce a parameter that shows
the fraction of successfully determined galaxies from the mock
catalog:
\begin{equation}
S_{\rm gal} = \frac {S_{\rm real}^{\rm gal} \cap S_{\rm rec}^{\rm
gal}}{S_{\rm real}^{\rm gal}}
\end{equation}
The denominator in this expression is equal to the number of the
same members in samples $S_{\rm real}^{\rm gal}$ and $S_{\rm
rec}^{\rm gal}$. This parameter shows what fraction of galaxies in
groups and clusters of the mock catalog were identified
automatically.

The second parameter characterizing the work of the program
algorithms for galaxies is the interloper fraction. This is the
fraction of all galaxies that were automatically grouped into
clusters, but which are field galaxies in the model catalog.
\begin{equation}
f_I = \frac {S_{\rm rec}^{\rm gal} \cap S_{\rm field}^{\rm
gal}}{S_{\rm rec}^{\rm gal}}
\end{equation}
Parameters $S_{\rm gal} $ and $f_I$ will also take values between
$0$ and $1$.

Evidently, it is impossible to have an ideal match between the
model catalog and the reconstructed groups and clusters. However,
we can optimize the work of the algorithms based on the parameters
presented above, using the quality parameters for the obtained
samples, determined as follows:

\begin{equation}
g_1=\sqrt{(1-c_1)^2+(1-p_1)^2},
\end{equation}
\begin{equation}
g_2 =\frac {c_2}{c_1} \frac {p_1}{p_2},
\end{equation}
\begin{equation}
g_3=\sqrt{(1- S_{\rm gal} )^2+{f_I}^2}.
\end{equation}

The parameter $g_1$ shows the balance between the completeness and
purity parameters of the derived sets of galaxy groups and
clusters, parameter $g_2$---that between the one-sided and
two-sided correspondence, and parameter $g_3$ is similar to $g_1$,
but, unlike $g_1$, which characterizes the detection of galaxy
groups with respect to the groups in the catalog, it shows the
relation between individual detected galaxies and groups. All
three parameters can take only values between~$0$ and~$1$. When
comparing the statistical parameters for the derived sets of
detected groups, one must try to minimize the parameters $g_1$ and
$g_3$, and maximize the parameter $g_2$.

Since the filtering method with adaptive kernel has the option to
vary its size, we computed for each of the considered apertures
the corresponding parameters of the detected cluster sample
quality in order to optimize the work of the algorithm. The best
parameters were obtained for the adaptive kernel size $8$. This
size was selected as optimal for further investigations.

For the Voronoi tesselation method, there are no parameters
similar to the method of adaptive kernel, which could be varied
within the algorithm. However, we can use different density
contrast values, which determine the membership of galaxies in
groups and clusters. The optimal density contrast based on the
quality parameters for the obtained group and cluster samples has
a value of two.

\section{VOID IDENTIFICATION}

\begin{figure}[bpt!!!]
\includegraphics[width=0.9\columnwidth]{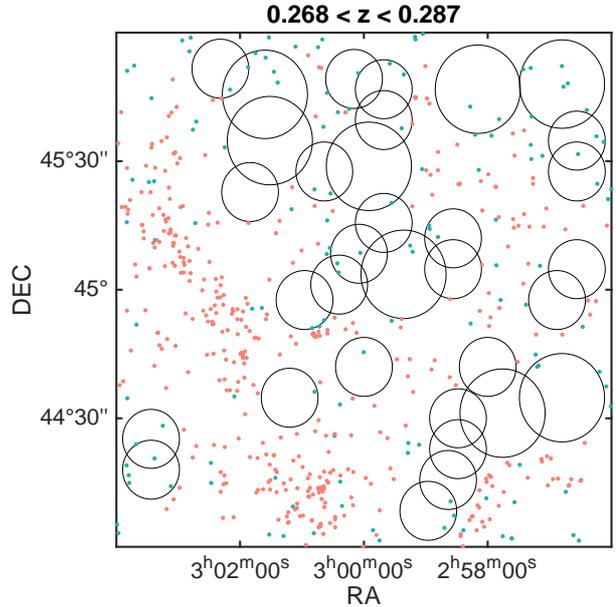}
\caption{A method of determining voids in the layers of the large
scale distribution of galaxies. The blue dots show galaxies for
which the computed density is 10 times smaller than average, and
the red dots show the remaining galaxies.} \label{fig_7:Grohovskaya_n_en}
\end{figure}

\begin{figure}[bpt!!!]
\includegraphics[width=\columnwidth]{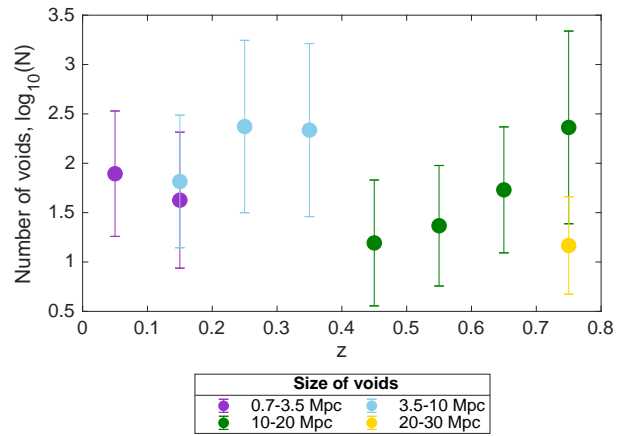}
\caption{Dependence of the number of detected voids of various
sizes from redshift. The dots correspond to the average void size,
the bars indicate the standard deviation from the average size of
the detected void at each $z$ for 10 sets of parameters from the
MICE simulation.} \label{fig_8:Grohovskaya_n_en}
\end{figure}

Isolating spatial regions of low density may also be useful for
further investigations based on the observed data for the physical
properties of galaxies as functions of the surrounding density.
The results obtained during the previous stage (density contrast
values for the galaxies of the large scale distribution) can be
used to determine galaxies located in voids. According to modern
notions, the density contrast of void galaxies is an order of
magnitude less than the average density of the distribution of
galaxies~\cite{Weygaert2011:Grohovskaya_n_en}.

The method used by us to determine voids is based on the algorithm
presented in~\cite{Patiri2006:Grohovskaya_n_en} for statistical
void analysis from the 2dFGRS~\cite{Madgwick2003:Grohovskaya_n_en}
survey. For determining voids in each layer we used a
$100\times100$ grid of points, which was superimposed onto each
layer separately. For each point we determined circles of maximum
diameter within which there are either no galaxies at all, or only
galaxies for which we determined in the previous step that the
density of their surroundings is one order of magnitude smaller
than the average density in the layer. The voids were restricted
from one side by the field boundaries, from the other~---by
galaxies for which the surrounding density was greater than the
density of the void galaxies. The step of the circle radii
variation was selected as $0.001$~radians, which corresponds to a
value from $0.297$ to $2.83$~Mpc depending from redshift. The step
size of the variation was chosen based on the optimal computing
time for close and distant redshifts.

We then selected the void-circles in such a way that the centers
of the selected circles should not lie within other identified
circles, and have the maximum radius among all those possible for
the given layer region. Additionally, with the redshift increase,
we imposed a restriction on the minimal volume of the detected
void.

An example of detecting voids for a thin slice of the three
dimensional galaxy distribution is presented in
Fig.~\ref{fig_7:Grohovskaya_n_en}. During the course of studying
the work of the algorithm for 10 sets of data from the MICE
simulation we discovered that the voids rarely have spherical
shape, being united into various structures (chains, etc.). For
the mock data of the MICE simulation we estimated the average
sizes of the identified voids and their number
(Fig.~\ref{fig_8:Grohovskaya_n_en}) as a function of redshift. As
is evident from the plot, mini-voids with sizes of
approximately~\mbox{$0.7$--$3.5$~Mpc} are dominant at the redshift
\mbox{$z \sim 0.1$--$0.2$}, and with an increase in the size of
the field we are able to detect voids with an average size of
about~\mbox{$10$--$20$~Mpc.}

\section{CONCLUSIONS}
We developed algorithms of multiparameter analysis of the large
scale distribution of galaxies in narrow slices in the entire
range of redshifts, including methods of determining clusters and
groups of galaxies by constructing density contrast maps, and also
determining voids. The results of using these algorithms
correspond to the visual estimate of the structures in slices, and
the average sizes of the detected clusters and voids correspond to
the current estimates with account of the size of the considered
field. A numerical estimation of the results of the methods of
density contrast map construction and identification of groups and
clusters was made for 10 data sets for galaxies and their clusters
from the MICE simulation, which allowed us to estimate the
completeness and purity of the sample, and also the fraction of
correctly identified galaxies as well as galaxies which were
misclassified as cluster galaxies. We can conclude that the
developed algorithms are a stable method of determining galaxy
group and cluster candidates as well as intergalactic void
candidates.

The method presented in this work will be used in the future to
study the large scale distribution of galaxies using the observed
data obtained with the 1 meter Schmidt telescope of the Byurakan
astrophysical observatory, and also to investigate the influence
of the surroundings on the physical properties of galaxies (mass,
luminosity, star formation rate).

\section{ACKNOWLEDGMENTS}
The authors are grateful to the referee for constructive feedback
on the initial text of the paper. This work has made use of
CosmoHub. CosmoHub has been developed by the Port d'Informacio
Cientifica (PIC), maintained through a collaboration of the
Institut de Fisica d'Altes Energies (IFAE) and the Centro de
Investigaciones Energeticas, Medioambientales y Tecnologicas
(CIEMAT), and was partially funded by the ``Plan Estatal de
Investigacion Cientifica y Tecnica y de Innovacion''\ program of
the Spanish government.

\section*{FUNDING}
This work was carried out with the financial support of the
Russian Science Foundation, project 17-12-01335 ``Ionized gas in
galactic discs and beyond the optical radius''.

\section*{CONFLICT OF INTEREST}
The authors declare no conflict of interest.


\end{document}